\documentclass[10pt]{article}
\usepackage[a4paper]{geometry}
\usepackage[utf8]{inputenc}
\usepackage[T1]{fontenc}
\usepackage{setspace,authblk}
\usepackage{amsmath,mathrsfs,amsfonts,amsthm}

\usepackage[english]{babel}

\usepackage{graphicx}

\usepackage{hyperref}

\title{\singlespace {Taxpayer deductions and the endogenous probability of tax penalisation}}

\author[]{Alex A.T. Rathke\thanks{NECCT/FEA-RP/USP, University of S\~ao Paulo. Avenida dos Bandeirantes 3900, 14040-905 Ribeir\~ao Preto, SP, Brazil. E-mail: \texttt{alex.rathke@alumni.usp.br}}}

\date{\today}

\theoremstyle{plain}

\newtheorem*{theorem*}{Theorem}

\newtheorem*{proposition*}{Proposition}

\newtheorem*{remark*}{Remark}

\newtheorem*{corollary*}{Corollary}

\begin{document}

\maketitle

\begin{abstract} 

We propose a parametric specification of the probability of tax penalisation faced by a taxpayer, based on the amount of deduction chosen by her to reduce total taxation. Comparative analyses lead to a closed-form solution for the optimum tax deduction, and provide the maximising conditions with respect to the probability parameters.

\end{abstract}


\section{Introduction} \label{Introduction}

Theoretical literature on tax evasion and optimal tax decisions predominantly follows the studies of \cite{allingham1972} and \cite{srinivasan1973}. In the standard model, the taxpayer aims to obtain a gain from under-reporting the true amount of taxable income, however she may be penalised with legal fines if the evasion is caught by tax authorities. Overall results show that, if the expected penalty cost is convex increasing, then there may be an equilibrium amount of non-reported income which maximises her expected utility.

Most studies assume that the probability of penalisation is either convex increasing or completely exogenous, which leads to the existence of the equilibrium. These assumptions imply analytical limitations. On the one hand, a fully-convex probability applies only to an upper-closed support, e.g. if the penalisation probability is a function of the total income, then it is not differentiable on the full support. On the other hand, an exogenous probability function limits the analysis of the taxpayer's behaviour \cite{alm2021}, as her perception of risk will not change if the amount of under-reported income changes.

This study derives a specification of the probability of penalisation, as a function of the amount of tax deduction chosen by the taxpayer to reduce her total taxation. We apply the properties of the traditional hazard function \cite{cox1972} to obtain an endogenous probability function with generalised parameters. For any probability function defined on the support $[0, \infty)$, we find that the optimal tax deduction exists if the mode of the probability distribution is not zero. Comparative analyses show how the probability parameters reflect the changes in tax enforcement policies. Our model applies directly to the standard tax evasion studies, where the penalisation probability is a function of the total income or the reported income.

\section{The model} \label{The model}

\subsection{Taxable income and optimal tax deduction} \label{Taxable income and optimal tax deduction}

Let a taxpayer obtain income $\pi(w)$ based on her own effort $w$, which must be taxed at a tax rate $t \in [0,1]$. Tax rules allow the taxpayer to deduce some amount $m \geq 0$ before paying taxes. Tax rules also require the taxpayer to report her income minus deductions to the tax authorities, for mandatory inspection. Assume the taxpayer cannot under-report $\pi$, and the after-tax income enjoyed by the taxpayer is $U(w) = \pi - t(\pi - m)$. 

The taxpayer analyses the tax rules and chooses the amount $m$ by self-assessment, however she has no information about how the tax authorities will evaluate $m$. If tax authorities interpret the deduction $m$ as abusive, they reclaim the payment of taxes over some excess deduction $\beta m$, $\beta \in [0,1]$, plus a penalty rate $z \geq 0$ over this amount to compensate the abuse \cite{allingham1972,srinivasan1973}.

Larger deductions are more likely to be scrutinised by tax authorities. For the event of penalisation by tax authorities $h \in \Omega$, define a non-negative random variable $X(h) : \Omega \rightarrow \mathbb{R}_{+}$. The probability of $X$ to be located within some interval $[0,m]$ is equal to the cumulative probability of $X$ at $m$, $P(X \leq m) = F_{X}(m) = F$, which is the same as the probability of the event of tax penalisation $h$ given the deduction $m$. The usual properties apply: $F(0) = 0$, $\partial F/ \partial m = f > 0$, $\lim_{m \to \infty} F = 1$. 

The expected after-tax income of the taxpayer is

\begin{equation} \label{expected after tax profit}
E(U) = \pi - t(\pi - m) - F \cdot [\beta mt(1 + z)] . \\
\end{equation}

Simplify the expression $\beta(1+z) = B$. Maximisation with respect to $m$ leads to the first and second-order conditions respectively equal to

\begin{equation} \label{first order m}
1 = (F + fm)B ; \\
\end{equation}

\begin{equation} \label{second order m}
\dfrac{\partial f}{\partial m} = f^{\prime} > - \dfrac{2f}{m} . \\
\end{equation}

Eq. \ref*{first order m} derives a positive equilibrium $\text{arg max}_{m} E(U) = m_{*} > 0$, since we have $1 > [F(0) + f(0) \cdot 0] = 0$ for any probability function $F$. The convexity condition in Eq. \ref*{second order m} shows that $m_{*}$ is a maximum point if the probability distribution $f$ is increasing at the left of the support $[0, \infty)$ such that for some amount $m$, we have $m_{*} \in (0,m] \subset [0, \infty) \rightarrow f^{\prime}(m_{*}) > 0 > -2f(m_{*})/m_{*} $.

\subsection{The probability of tax penalisation and comparative statics} \label{The probability of tax penalisation and comparative statics}

We derive a specification of $F$ with respect to the endogenous variable $m$ based on a survival process.

The taxpayer chooses the amount of tax deduction $m > 0$, producing a probability of tax penalisation equal to $F_{X}(m) > 0$. Assume now that the taxpayer may increase the deduction to some amount $m + \delta$, $\delta > 0$, which reduces the total taxes by $\delta t$, however it increases the probability of penalisation. Given an initial deduction $m$, the increase in the probability of penalisation produced by $\delta$ is equal to

\begin{equation} \label{prob delta}
P(m \leq X \leq m + \delta | X \geq m) = \dfrac{F_{X}(m + \delta) - F_{X}(m)}{1 - F_{X}(m)} . \\
\end{equation}

Dividing Eq. \ref*{prob delta} by $\delta$ and taking the limit $\delta \rightarrow 0$, we obtain

\begin{equation} \label{hazard function}
\lim_{\delta \to 0} \dfrac{F_{X}(m + \delta) - F_{X}(m)}{\delta \cdot [1 - F_{X}(m)]} = \dfrac{f_{X}(m)}{1 - F_{X}(m)} = \lambda , \\
\end{equation}

\noindent which defines the traditional hazard function $\lambda(m)$ \cite{cox1972} satisfying $m \geq 0 \rightarrow \lambda \geq 0$, $\int_{0}^{\infty} \lambda dm = \infty$. Integrating both sides of Eq. \ref*{hazard function} on the interval $[0,m]$ and solving for $F_{X}(m) = F$, we obtain the probability of tax penalisation equal to

\begin{equation} \label{prob}
F = 1 - e^{- \int_{0}^{m} \lambda dy} . \\
\end{equation}

We assume that the hazard of penalisation $\lambda$ is a monotone increasing function of $m$, since tax authorities are stricter against larger tax deductions \cite{alm2021}. Therefore, conventional Laplace transform allows a parametric representation; i.e. we may parametrise $\lambda = Am^{k-1}$, where $A > 0$ and $k > 0$ are general scale and shape parameters respectively\footnote{Formally, any completely monotone function $g : (0, \infty) \rightarrow \mathbb{R}_{+}$ may be represented as a mixture of exponentials, see \cite{mchugh1975}.}, for the probability of tax penalisation becomes

\begin{equation} \label{prob param}
F = 1 - e^{-\frac{A}{k}m^{k}} . \\
\end{equation}



Substituting Eq. \ref*{prob param} in Eq. \ref*{first order m} and Eq. \ref*{second order m}, we obtain the optimal tax deduction $m_{*}$ and the maximising conditions in parametric form, for we have

\begin{equation} \label{maximum m}
\begin{array}{rl}
0 < m_{*} &= \left( \dfrac{1-W k}{A} \right)^{1/k} < \left( \dfrac{1 + k}{A} \right)^{1/k} ; \\
\\
-1 < W &= W_{0} \left[ \left(1 - \dfrac{1}{B} \right) \cdot \dfrac{e^{1/k}}{k} \right] < \dfrac{1}{k} ; \\
\\
- \dfrac{1}{e} &< \left(1 - \dfrac{1}{B} \right) \cdot \dfrac{e^{1/k}}{k} ; \\
\\
0 < B < 1 &\rightarrow 1 < W_{0} \left( \dfrac{B}{e(1 - B)} \right)^{-1} < k , \\
\end{array}
\end{equation}

\noindent where $W_{0} : \mathbb{R} \rightarrow [-1, \infty)$ is the principal real-valued branch of the Lambert $W$-function \cite{iacono2017}, which provides an analytical solution for $m_{*}$. The closed-form expression for $m_{*}$ is straightforward. The inequalities in Eq. \ref*{maximum m} with respect to $W$, $k$, $B$ derive from the second-order condition for $m_{*}$ to be a maximum point\footnote{$k > 1$ means that the probability function $F$ is S-shaped with a convex region at the left of the support $[0, \infty)$, therefore the optimal point $m_{*}$ exists for some interval $m_{*} \in (0,m]$, $f^{\prime}(m_{*}) > 0$. If $k \leq 1$, the probability $F$ is concave on the full support, for the second-order condition in Eq. \ref*{second order m} is not satisfied.}.

Assume the conditions in Eq. \ref*{maximum m} are satisfied. Comparative analyses regarding $m_{*}$ provide the following:

\begin{equation} \label{dmdA}
\dfrac{\partial m_{*}}{\partial A} = - \dfrac{m_{*}}{Ak} < 0 ; \\
\end{equation}
\begin{equation} \label{dmdB}
\dfrac{\partial m_{*}}{\partial B} = - \dfrac{m_{*}^{1-k}}{AB^2 \cdot (1 - 1/B) } \cdot \dfrac{W}{W + 1} < 0 ; \\
\end{equation}
\begin{equation} \label{dmdk}
\dfrac{\partial m_{*}}{\partial k} = \dfrac{m_{*}}{k} \cdot \left( \dfrac{W - ( 1 + 1/k ) \frac{W}{W + 1} }{Wk - 1} - log(m_{*}) \right) . \\
\end{equation}

The negative effects from variables $A$ and $B$ are intuitive, since $A$ is an inverse scale of $m_{*}$, while $B$ increases after increase of either the reclaim value $\beta$ or the penalty rate $z$ \cite{allingham1972,srinivasan1973}. The impact of the shape parameter $k$ in Eq. \ref*{dmdk} is more intricate. Close inspection shows that the derivative $\partial m_{*} / \partial k$ depends on the variation of $W \in (-1,1/k)$ within its range, such that $W \geq 0 \rightarrow \partial m_{*} / \partial k > 0$, $\lim_{W \to -1} \partial m_{*} / \partial k = -\infty$, $\lim_{W \to 1/k} \partial m_{*} / \partial k = 0_{+}$. Since Eq. \ref*{dmdk} is continuous, it means that $\partial m_{*} / \partial k$ is inverse-U shaped with a positive maximum, and there is a value $W$ for which Eq. \ref*{dmdk} is zero.

\section{Discussion} \label{Discussion}

In our analysis, the probability $F$ is fully specified by the parameters $A$ and $k$\footnote{Eq. \ref*{prob param} becomes a version of the Weibull distribution.}, which models the expected behaviour of tax authorities given the amount of tax deduction $m$. Higher tax enforcement implies a higher probability of penalisation, so the curve $F$ moves to the left of the support, which implies a larger $A$ or lower $k$. Lower tax enforcement implies the opposite effect on parameters. Eq. \ref*{prob param} applies directly to the usual enforcement policies including audit rules and fines \cite{alm2021}. Variations on the hazard function $\lambda$ may reflect different assumptions or more detailed parametrisations, which derive the probability function $F$ by direct manipulation within Eq. \ref*{prob}.

Some extensions are immediate. For the standard model in tax literature, $m$ represents the amount of evaded income, and $F$ may be a function of the reported income $\pi - m \geq 0$ \cite{allingham1972}, with analogous specification. If $m$ is limited to a closed interval, say $m \in [0, \pi]$, $F(\pi) = 1$, then Eq. \ref*{prob param} simplifies to

\begin{equation} \label{prob param closed support}
F(m \in [0, \pi]) = 1 - \left( 1 - \dfrac{m^{k}}{\pi} \right)^{\frac{A \pi}{k}} , \\
\end{equation}

\noindent therefore the probability $F$ becomes a function of the tax deduction $m$ and the total income $\pi$ \cite{srinivasan1973}. For analytical simulations or estimation of the optimal point $m_{*}$, the Lambert W-function in Eq. \ref*{maximum m} adopts several approximations, see \cite{iacono2017}.



\end{document}